\documentclass[onecolumn,superscriptdaddress,notitlepage,showpacs]{revtex4-1}
\usepackage{amsmath}
\usepackage{graphicx}
\usepackage{bm}
\usepackage{amsbsy}
\usepackage{amsfonts}
\usepackage{amsthm}
\usepackage[dvips]{color}

\newcommand{\diff}[2]{\frac{\partial #1}{\partial #2}}

\newcommand{\gap}{\gamma}

\newcommand{\bra}[1]{\langle #1|}
\newcommand{\ket}[1]{|#1\rangle}
\newcommand{\braket}[2]{\langle #1|#2\rangle}
\newcommand{\ketbra}[2]{|#1\rangle\!\langle #2|}

\begin{document}
\title{Improved Error-Scaling for Adiabatic Quantum State Transfer}
\author{Nathan Wiebe$^{1,2}$ and Nathan S. Babcock$^1$}
\affiliation{$^1$ Institute for Quantum Information Science, University of Calgary, Alberta, Canada\\$^2$Institute for Quantum Computing, University of Waterloo, Ontario, Canada}

\begin{abstract}
We present a new technique that improves the scaling of the error in the adiabatic approximation with respect to the evolution duration, thereby permitting faster transfer at a fixed error tolerance.
Our method is conceptually different from previously proposed techniques: it exploits a commonly overlooked phase interference effect that occurs predictably at specific evolution times, suppressing transitions away from the adiabatically transferred eigenstate.  Our method can be used in concert with existing adiabatic optimization techniques, such as local adiabatic evolutions or boundary cancellation methods. We perform a full error analysis of our phase interference method along with existing boundary cancellation techniques and show a tradeoff between error-scaling and experimental precision.  We illustrate these findings using two examples, showing improved error-scaling for an adiabatic search algorithm and a tunable two-qubit quantum logic gate.
\end{abstract}
\maketitle


The adiabatic approximation underpins many important present-day and future applications, such as stimulated rapid adiabatic passage (STIRAP) \cite{oreg:stirap,kuklinski:stirap},
coherent control of chemical reactions \cite{Shapiro:2006}, and quantum information processing (QIP) \cite{averin:qcomp, Babcock:2008}. This approximation asserts that a system will remain in an instantaneous eigenstate of a time-varying Hamiltonian if the time-variation happens slowly enough. Errors in this approximation correspond to transitions away from the instantaneous (``adiabatically transferred'') eigenstate. For high-performance applications, it is not always practical to minimize errors by slowing things down. Ambitious future technologies, such as quantum computing devices, will demand simultaneous maximization of both accuracy and speed.



In this paper, we investigate a phase cancellation effect that appears during an adiabatic evolution and can be exploited to  polynomially reduce the probability of a given transition at fixed maximum evolution time.  This can lead to speed increases at fixed error probability.  Unlike alternative methods that obtain improvements by modifying the adiabatic path~\cite{roland:localad, rezakhani:geodesic}, our technique chooses the evolution time so that destructive interference suppresses the transition.  Furthermore, this phase cancellation effect can be exploited to improve existing adiabatic error reduction strategies such as local adiabatic evolutions or boundary cancellation methods.  We provide an error analysis of our method and conclude that the accuracy improvements come at the price of increasingly precise knowledge of the time-dependent Hamiltonian; this implies that accuracy is an important and quantifiable resource for quantum protocols utilizing adiabatic passage.


\section{Adiabatic Approximation}
Following previous authors~\cite{mackenzie:adiabatic,ruskai:adiabatic}, we define the error $\mathcal{E}$ to be the component of the post-evolution state vector that is orthogonal to the state intended for adiabatic transfer. For convenience, we represent all mathematical terms as explicit functions of the ``reduced time'' $s(t)=t/T$, where $t$ is the time, $T$ is the total evolution duration, and $0\le s\le 1$. This parameterization leaves the form of the Hamiltonian $\mathcal{H}(s)$ unchanged as $T$ varies.

In many circumstances, the following criterion adequately estimates the magnitude of the total error $\mathcal{E}$ at time $t = T$ for a given Hamiltonian:
\begin{equation}
\|\mathcal{E}\| \lesssim\frac{1}{T}\max_s\frac{\|\frac{\mathrm{d}}{\mathrm{d}s} \mathcal{H}(s) \|}{\min_{\nu} |E_\nu(s)-E_0(s)|^2},\label{eq:standardcriterion}
\end{equation}
where $E_\nu(s)$ ($\nu\neq0$) is the instantaneous energy of the $\nu^\text{th}$ eigenstate of the Hamiltonian $\mathcal{H}(s)$ and $E_0(s)$ is the energy of the eigenstate being transferred (usually the ground state)~\cite{farhi:adiabatic,roland:localad}. We use the convention $\hbar=1$.

Although eq.~\eqref{eq:standardcriterion} provides an expedient heuristic for estimating the accuracy of adiabatic passage, it is (in general) neither necessary nor sufficient to bound the fidelity of adiabatic state transfer \cite{marzlin:counter, teufel:adiabatic}. This equivocality opens the possibility of a modest allocation of resources being used to enable significantly improved error-scaling.

One method of improving the fidelity of adiabatic transfer is via the use of a ``local adiabatic'' evolution~\cite{vandam:adiabaticpower,roland:localad,rezakhani:geodesic}.  The idea behind the local adiabatic approximation is to tailor variation of $\mathcal{H}$ with respect to $s$ to minimize the instantaneous nonadiabatic transition rate ${\|\diff{}{s} \mathcal{H}(s)\|}/{\min_\nu\!|E_\nu(s)-E_0(s)|^2}$.  Local adiabatic methods have lead to substantial improvements in the asymptotic error-scaling $\mathcal{E}$ with respect to the Hilbert space dimension $N$~\cite{vandam:adiabaticpower,roland:localad,rezakhani:geodesic}; however, they do not  improve the scaling of the error with $T$.

Recently, methods were developed for improving the scaling of $\mathcal{E}$ with $T$ from order $\mathcal{O}(1/T)$ to $\mathcal{O}(1/T^{m+1})$ by setting the first $m$ derivatives of the Hamiltonian to zero at the beginning and end of the evolution~\cite{rezakhani:adiabaticexponential,lidar:exponentialtime}.  Error reduction techniques employing this result are collectively referred to as ``boundary cancellation methods.''   Boundary cancellation methods have two main drawbacks: first, they assume that the first $m$ derivatives of $\mathcal{H}(s)$ are exactly zero, leaving it unclear whether they are robust against small variations in the derivatives of the Hamiltonian; second, in the regime of short $T$ these methods can have error-scaling that is inferior to the trivial case wherein no boundary cancellation technique is applied (i.e., $m=0$).  Our work addresses these problems: we first provide an analysis of the sensitivity of boundary cancellation methods to small variations in the values of the first $m$ derivatives of $\mathcal{H}(s)$; we then show that phase interference can be used to further reduce errors without increasing $m$, providing better error-scaling for short $T$.

\section{Main Result}

We present a new technique for quadratically suppressing the probability of a particular nonadiabatic transition during adiabatic passage. It works by exploiting a phase interference effect that appears in adiabatic systems with Hamiltonians obeying a simple symmetry. This effect can be exploited in a realistic class of time-dependent Hamiltonians that includes many adiabatic algorithms and transport protocols, as well as any Hamiltonian obeying $\mathcal{H}(0)=\mathcal{H}(1)$. 

Consider a time-dependent Hamiltonian $\mathcal{H}(s)$ acting on an $N$-dimensional Hilbert space spanned by the instantaneous energy eigenvectors $\ket{\nu(s)}$ where $\nu=0,1,\ldots, N-1$.  We define $|0(s)\rangle$ to be the state intended for adiabatic passage.  We use the notation $\mathcal{H}^{(p)}(s)=(\diff{}{x})^p \mathcal{H}(x)|_s$. In Section~\ref{sec:theory} we will show that errors in adiabatic passage can be reduced for Hamiltonians obeying the boundary symmetry condition,
\begin{equation}
\frac{\bra{\nu(1)}{\mathcal{H}}^{(m+1)}(1) \ket{0(1)}}{\big(E_\nu(1)-E_0(1)\big)^{m+2}}=\left(\frac{\bra{\nu(0)}{\mathcal{H}}^{(m+1)}(0) \ket{0(0)}}{\big(E_\nu(0)-E_0(0)\big)^{m+2}}\right)e^{-i\theta},\label{eq:condition}
\end{equation}
where $\theta$ is an arbitrary phase factor, and $m$ is the number of derivatives of $\mathcal{H}(s)$ that are zero at the boundaries $s=0,1$ (e.g., if $m=2$ then the first and second derivatives of $\mathcal{H}(s)$ are zero at the boundaries, whereas if $m=0$ then none are zero on the boundary). In practice, any time-dependent Hamiltonian may be adapted to satisfy eq.~\eqref{eq:condition}, simply by adjusting its rate of change in $s$ at the boundaries.  If eq.~\eqref{eq:condition} is not exactly satisfied then the phase interference effect will still reduce errors, but it will not necessarily improve the asymptotic error-scaling with $T$.

Our method can be used in conjunction with existing boundary cancellation methods to produce even greater improvements in the asymptotic error-scaling with $T$. Amplitudes of the transitions $\ket{0(0)}\rightarrow\ket{\nu(1)}$ are reduced from the order $\mathcal{O}(T^{-m-1})$ estimates given in refs.~\cite{rezakhani:adiabaticexponential,lidar:exponentialtime} to order $\mathcal{O}(T^{-m-2})$ at the discrete set of times $T=T_{n,\nu}$, where $n$ is an even integer and
\begin{equation}
T_{n,\nu}=\frac{n\pi-\theta}{\int_0^{1} [E_{\nu}(s)-E_{0}(s)]\mathrm{d}s}.\label{eq:searchDestruct}
\end{equation}
This can lead to polynomial reductions in the \emph{overall} error if $\|\mathcal{E}\|$ is dominated by a small number of transitions.

We refer to boundary cancellation methods that are augmented by our scheme to produce order $\mathcal{O}(T^{-m-2})$ error-scaling as ``augmentented boundary cancellation methods.''
In Section~\ref{sec:error1}, we will analyze the error robustness of our augmented boundary cancellation method along with the original schemes laid out in refs.~\cite{rezakhani:adiabaticexponential,lidar:exponentialtime}. We show that performance improvements are derived from accurate knowledge of the system's eigenspectrum $\{E_\nu\}$, its total evolution time, and the derivatives of its Hamiltonian, and we provide quantitative error-bounds on these quantities. We provide numerical examples that verify the predictions of our theory in Sections~\ref{sec:search} and~\ref{sec:gate}.


\section{Theory\label{sec:theory}}

We will break our discussion of the theory of our method into two parts.  First, we discuss the special case for which $m=0$.  This 
simple case is conceptually distinct from existing boundary cancellation techniques, which require $m > 0$ to produce improvements over eq.~\eqref{eq:standardcriterion}. We then discuss the more general case in which $m>0$.

To obtain our results, it is \emph{not} necessary to assume that the instantaneous eigenvalues satisfy the ordering condition $E_0(s)\!<\!E_1(s)\!<\!\ldots\!<\! E_{N-1}(s)$. We do however require that $E_0(s) \neq E_\nu(s)~\forall~\nu>0$, unless transitions between $\ket{0(s)}$ and $\ket{\nu(s)}$ are strictly forbidden by $\mathcal{H}(s)$.  For convenience, we also assume that the phases of the instantaneous eigenvectors are chosen such that $\braket{\dot\nu(s)}{\nu(s)}=0$.  This choice does not affect the quantum dynamics, but it simplifies the analysis of the error.  {W}e also assume that the Hamiltonian is differentiable $m+2$ times and that each derivative is bounded for all $T$.  These last restrictions are put in place order to prevent issues that arise for Hamiltonians resembling that of the Marzlin--Sanders counterexample~\cite{marzlin:counter,cheung:adiabatic}.

Given the above assumptions, the error in the adiabatic approximation $\mathcal{E}$ for a Hamiltonian evolution acting on an $N$-dimensional Hilbert space is given by
\begin{equation}
\mathcal{E}=\sum_{\nu=1}^{N-1}\mathcal{E}_\nu e^{-iT\!\int_0^1 E_\nu(s)\mathrm{d}s}\ket{\nu(1)}+\mathcal{O}(T^{-m-2})\label{eq:error}.
\end{equation}
We know from previous work that $\mathcal{E}_\nu\in \mathcal{O}(T^{-m-1})$~\cite{rezakhani:adiabaticexponential,lidar:exponentialtime}, and asymptotically tight expressions are known for $\mathcal{E}_\nu$ in the $m=0$ case~\cite{mackenzie:adiabatic,cheung:adiabatic}. We therefore begin with this case to illustrate how our phase interference effect can be utilized.  
Given that $m=0$, the form of $\mathcal{E}_\nu$ reduces to
\begin{equation}
\mathcal{E}_\nu =\left.\frac{\bra{\nu(s)}\dot{\mathcal{H}}(s) \ket{0(s)}e^{-iT\int_0^s \gap_{\nu}(\xi)\mathrm{d}\xi}}{-iT\gap_{\nu}^2(s)} \right|_{s=0}^1, \label{eq:firstorder1}
\end{equation}
and where $\gap_{\nu}(s)\!=\!E_0(s)-E_\nu(s)$. If we choose $\mathcal{H}(s)$ to obey~\eqref{eq:condition}, then the absolute value of eq.~\eqref{eq:firstorder1} reduces to
\begin{equation}
|\mathcal{E}_\nu|=\left|\frac{\bra{\nu(s)}\dot{\mathcal{H}}(s) \ket{0(s)}}{T\gap_{\nu}^2(0)}\!\left(e^{-i\left(\theta+T\int_0^1 \gap_{\nu}(s)\mathrm{d}s \right)}\!-\!1\right)\right|\label{eq:firstorder2}.
\end{equation}


Eq.~\eqref{eq:firstorder2} has extrema at $T=T_{n,\nu}$. It is maximized when $n$ is odd and vanishes when $n$ is even.  Thus, when $T=T_{n,\nu}$ (even $n$), phase interference causes the magnitude of the $\nu^\text{th}$ component of $\mathcal{E}$ to be quadratically reduced from $\mathcal{O}(T^{-1})$ to $\mathcal{O}(T^{-2})$.



If $m>0$ then the phase interference effect also suppresses probability of excitation to $\ket{\nu(1)}$ at $T=T_{n,\nu}$ for any even integer $n>0$, but this effect does not directly follow from existing results.  We show in Appendix~\ref{appendix:1} using a perturbative expansion (similar in reasoning to that of refs.~\cite{rezakhani:adiabaticexponential,lidar:exponentialtime}) that if the first $m$ derivatives of $\mathcal H (s)$ are zero at the boundaries $s=0,\!1$ then
\begin{align}
|\mathcal{E}_\nu|=\left|\left. \frac{\bra{\nu(s)}\mathcal{H}^{(m+1)}(s)\ket{0(s)}e^{-i\int_0^s\gap_{\nu}(\xi)\mathrm{d}\xi T}}{T^{m+1}\gap_{\nu}^{m+2}(s)}\right|_{s=0}^1\right|.\label{eq:generalboundary}
\end{align}

Similar to eq.~\eqref{eq:firstorder1}, eq.~\eqref{eq:generalboundary} reveals an adiabatic phase interference effect also that suppresses the error at certain times.  This suppression occurs when
\begin{equation}
\frac{\bra{\nu(1)}\mathcal{H}^{(m+1)}(1)\ket{0(1)}e^{-iT\int_0^1\gap_{\nu}(s)\mathrm{d}s}}{\gap_{\nu}(1)^{m+2}}=\frac{\bra{\nu(0)}\mathcal{H}^{(m+1)}(0)\ket{0(0)}}{\gap_{\nu}(0)^{m+2}},
\end{equation}
implying that adiabatic phase interference effects reduce the order of transition amplitude $\mathcal{E}_\nu$  from $\mathcal{O}(T^{-m-1})$ to $\mathcal{O}(T^{-m-2})$ when $T=T_{n,\nu}$ for even $n$.

\section{Tolerances}\label{sec:error1}

Limits on the precision of physical apparatus prevent perfect phase cancellation in realistic applications. 
Errors can result from imperfect modelling of the Hamiltonian, inexact calculations of the gap integrals, or inaccuracies in the timing or control apparatus. It is therefore necessary to address the impact of empirical imperfections on the feasibility of augmented boundary cancellation methods and determine when they methods can be experimentally realized.

``Symmetry errors'' occur when the timing symmetry condition~\eqref{eq:condition} is not precisely satisfied:
\begin{equation}
\Delta S_\nu=\left|\frac{\bra{\nu(1)}\mathcal{H}^{(m+1)}(1)\ket{0(1)}}{\gap_{\nu}(1)^{m+2}}-\frac{\bra{\nu(0)}\mathcal{H}^{(m+1)}(0)\ket{0(0)}}{\gap_{\nu}(0)^{m+2}}e^{-i\theta}\right| > 0.\label{eq:deltabetadef}
\end{equation}
Comparing eq.~\eqref{eq:deltabetadef} with eq.~\eqref{eq:generalboundary}, we find that the contributions to $\mathcal{E}_\nu$ due symmetry errors are of order $\mathcal{O}(T^{-m-2})$ so long as $\Delta S_\nu \in \mathcal{O}(T^{-1})$. 

``Gap errors'' occur when inaccuracies in the estimate of the gap integral leave condition \eqref{eq:searchDestruct} unsatisfied:
\begin{equation}
\Delta G_\nu=\left|\int_0^{1} \gamma_{\nu}(\xi)\mathrm{d}\xi - \frac{n\pi-\theta}{T_{n,\nu}}\right| > 0.
\label{eq:deltagammadef}
\end{equation}
Expanding eq.~\eqref{eq:firstorder2} in powers of $\Delta G_\nu$, we find that the contributions to $\mathcal{E}_\nu$ due to gap errors are of order $\mathcal{O}(T^{-m-2})$ if $\Delta G_\nu\in\mathcal{O}(T^{-2})$.

``Timing errors'' occur when the actual evolution time $T$ differs from the ideal evolution time $T_{n,\nu}$:
\begin{equation}
\Delta T_{n,\nu}=\left|T_{n,\nu}-T\right| > 0. \label{eq:deltaGammadef}
\end{equation}
Expanding eq.~\eqref{eq:firstorder2} in powers of $\Delta T_{n,\nu}$, we find that the contributions to $\mathcal{E}_\nu$ due to timing errors are of order $\mathcal{O}(T^{-m-2})$ if $\Delta T_{n,\nu}\in \mathcal{O}(T^{-1})$.

``Derivative errors'' can also occur wherein one or more of the derivatives of the Hamiltonian that is assumed
to be zero is not:
\begin{equation}
\Delta \mathcal{H}^{(p)} =\max_{s=0,1} \|\mathcal{H}^{(p)}(s)\| >0,
\end{equation}
for  $p=1,\ldots,m$.
Such errors do not affect the error-scaling if for all such $p$,
\begin{equation}
{\Delta \mathcal{H}^{(p)}}\in \mathcal{O}(1/T^{m+2-p}).\label{eq:derivErrorTol}
\end{equation}

In other words, given that the first $m$ derivatives of $\mathcal{H}$ are \emph{approximately} zero at the boundaries, the uncertainty in each derivative must shrink polynomially as $T$ increases in order to achieve the full promise of an augmented boundary cancellation method. The proof that this criteria is sufficient is not simple: it requires a high-order perturbative analysis of the error in the adiabatic approximation.  Details are provided in Appendix~\ref{appendix:2}.


If $m$ is a constant, then it follows that augmented boundary cancellation methods are error robust in the sense that their error tolerances scale polynomially with $T^{-1}$.  This is not problematic for numerical studies because additional precision can be provided at poly-logarithmic cost. However, experimental errors cannot always be so conveniently reduced, and boundary cancellation techniques that use a large value of $m$ may be impractical. The situation is even worse if exponential error-scaling is required, which can be obtained if $m \in \Theta( T/\log T)$. In such circumstances the tolerances $\mathcal{H}^{(p)}(s)$ decrease exponentially with $T$ and therefore boundary cancellation methods are not error robust. This implies that boundary cancellation techniques (augmented or not) cannot in practice achieve exponential scaling without exceedingly precise knowledge of the derivatives of the Hamiltonian at the boundaries.  The $m=0$ method may therefore be more experimentally relevant than its higher-order brethren, because of its minimal precision requirements and its superior scaling for modestly short $T$.

As the performance improvements provided by boundary cancellation methods come at the price of increasingly accurate information about the Hamiltonian and the evolution time, such information may be viewed as a computational resource for protocols utilizing quantum adiabatic passage.
This suggests that current analyses \cite{AQC=circuit-based} of the resources required for generic adiabatic quantum computing may be incomplete. We illustrate this subtlety in Section~\ref{sec:search} by showing how to quadratically improve the total error-scaling $||\mathcal{E}||$ of an already ``optimal'' quantum algorithm.

\section{Search Hamiltonians}\label{sec:search}

Adiabatic quantum computing (AQC) algorithms are natural candidates for error suppression by our technique.  To demonstrate, we examine an algorithm that adiabatically transforms an initial guessed state into the sought state of a search problem~\cite{farhi:adiabatic}.  The Hamiltonian for this algorithm is

\begin{equation}
\mathcal{H}(s)=I-(1-\phi(s))\ket{+^{\otimes{n}}}\bra{+^{\otimes{n}}}-\phi(s)\ketbra{0^{\otimes{n}}}{0^{\otimes{n}}}, \label{eq:searchHam}
\end{equation}
where $\ket{+}=(\ket{0}+\ket{1})/\sqrt{2}$, $\ket{0^{\otimes n}}$ is the state that the algorithm seeks, and $\phi:[0,1]\mapsto[0,1]$ obeys $\phi(0)=0$ and $\phi(1)=1$.

\begin{figure*}[t!]
\begin{minipage}[t]{0.45\linewidth}
\center{\includegraphics[width=\linewidth]{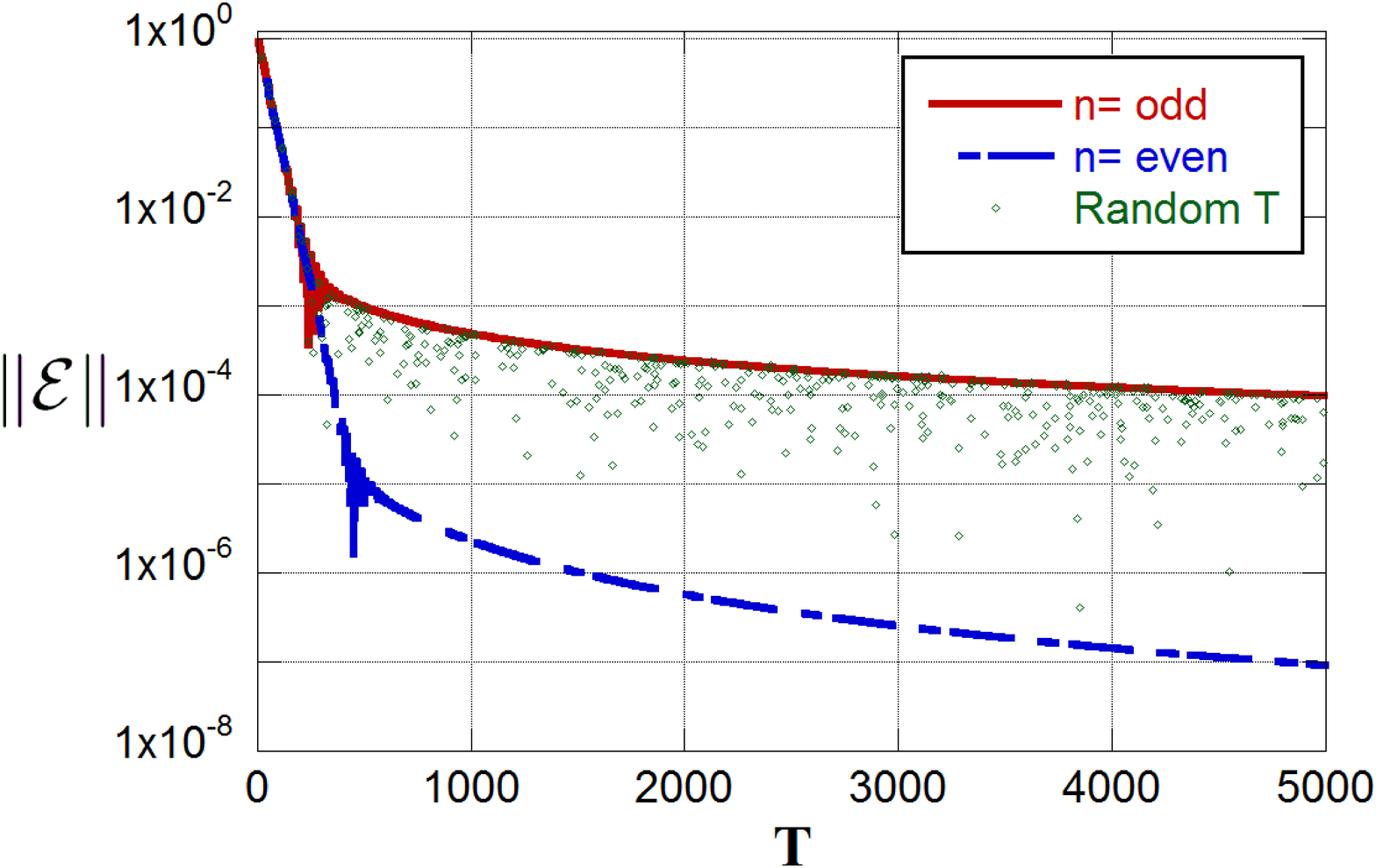}}
\caption{Final error amplitude $|\mathcal{E}|$ as a function of $T$ for the Search Hamiltonian \eqref{eq:searchHam} using $N=16$ and $\phi(s)=s$. \label{fig:adiabaticsubsequence}}
\end{minipage}
\hspace{1cm}
\begin{minipage}[t]{0.45\linewidth}
\center{\includegraphics[width=\linewidth]{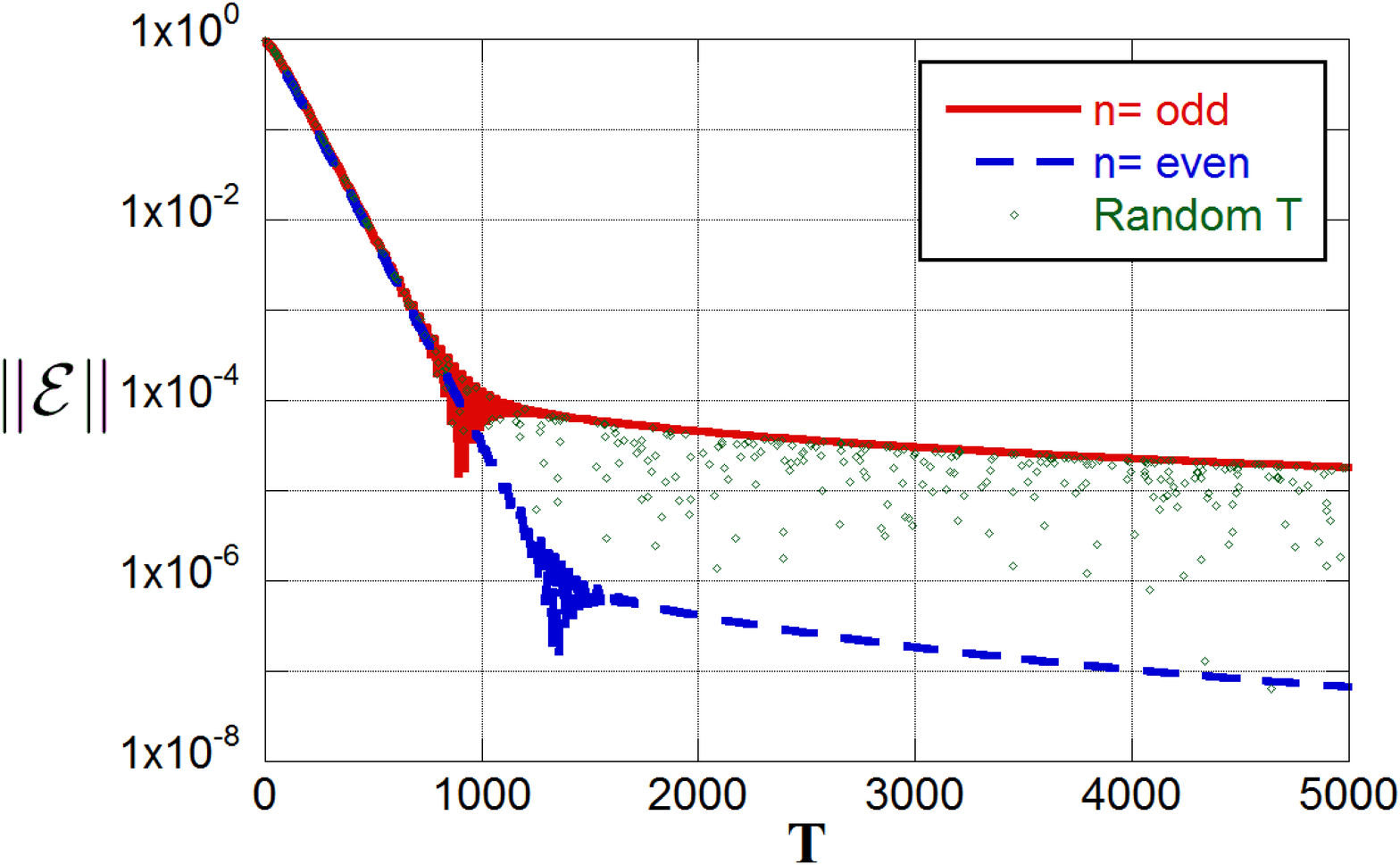}}
\caption{Final error amplitude $|\mathcal{E}|$ as a function of $T$ for the Search Hamiltonian \eqref{eq:searchHam} using $N=16$ and eq.~\eqref{eq:localevolution}.\label{fig:localadiabaticsubsequence}}
\end{minipage}
\end{figure*}

Two common choices for $\phi(s)$~\cite{roland:localad,vandam:adiabaticpower,farhi:adiabaticpaths} are $\phi(s)=s$ and
\begin{equation}
\phi(s)=\frac{\sqrt{N-1}-\tan\big[\arctan(\sqrt{N-1})(1-2s)\big]}{2\sqrt{N-1}}\label{eq:localevolution}.
\end{equation}
The latter choice~\eqref{eq:localevolution} is said to generate a ``local'' adiabatic evolution \cite{vandam:adiabaticpower,roland:localad}.
In each case, the (dimensionless) energy gap is
\begin{equation}
\gap_{1}(s)=\sqrt{1-4\left(1-\frac{1}{N} \right)\phi(s)(1-\phi(s))},
\end{equation}
where $\ket{0(s)}$ is the ground state of eq.~\eqref{eq:searchHam} and $\ket{1(s)}$ is the only other eigenstate that is coupled to $\ket{0(s)}$~\cite{roland:adsim}. From the eigenvectors of $\mathcal{H}(s)$, it is straightforward to verify that both forms of $\phi(s)$ given above satisfy eq.~\eqref{eq:condition} with $m=0$.


Figs.~\ref{fig:adiabaticsubsequence} and~\ref{fig:localadiabaticsubsequence}
show that the choice $T=T_{n,\nu}$ (even $n$) produces quadratic improvements in the scaling of $\|\mathcal{E}\|$ for both $\phi(s)=s$ and eq.~\eqref{eq:localevolution} at large $T$.  For odd values of $n$, the error is maximized, as expected.  It is apparent that randomly selected times are extremely unlikely to exhibit maximum phase cancellation. 
Figs.~\ref{fig:adiabaticsubsequence} and~\ref{fig:localadiabaticsubsequence} also suggest a second benefit of our technique:  existing boundary cancellation methods \cite{rezakhani:adiabaticexponential,lidar:exponentialtime} can improve the performance of adiabatic algorithms in the limit of large $T$, but these improvements come at the price of inferior error-scaling for small $T$, as seen in Fig.~\ref{fig:boundarycancel} of ref.~\cite{rezakhani:adiabaticexponential}.  The results shown here in Figs.~\ref{fig:adiabaticsubsequence} and~\ref{fig:localadiabaticsubsequence} exhibit no such tradeoff.

Figs.~\ref{fig:adiabaticsubsequence} and~\ref{fig:localadiabaticsubsequence} also shed light on the nature of the complexity of adiabatic algorithms.  Several previous studies have taken the complexity of an adiabatic algorithm to be given by the evolution time required for the error predicted by eq.~\eqref{eq:standardcriterion} to fall within a specified tolerance~\cite{farhi:adiabatic,vandam:adiabaticpower,roland:localad}. In the case of the local adiabatic evolution, this time scales as $\mathcal{O}(\sqrt{N})$, which is known to be optimal~\cite{roland:localad,vandam:adiabaticpower}.  Fig.~\ref{fig:localadiabaticsubsequence} show that this error can still be quadratically reduced by eliminating the $\mathcal{O}(T^{-1})$ contributions to it. These results do not violate quantum lower bounds because the time required for the $\mathcal{O}(1/T)$ to become dominant still scales as $\mathcal{O}(\sqrt{N})$~\cite{rezakhani:adiabaticexponential}.  Therefore even an exponential improvement in the subsequent adiabatic regime would not violate quantum lower bounds. Paradoxically, these results suggest that the complexity of adiabatic algorithms may be dictated by the physics of the sudden approximation rather than the adiabatic approximation.

We demonstrate our generalized $m>0$ technique in Fig.~\ref{fig:boundarycancel}, where we plot $|\mathcal{E}_\nu|$ as a function of the total evolution time for
a search Hamiltonian with $\phi(s)$ taken to be
\begin{equation}
\phi(s)=\frac{\int_0^s x^m(1-x)^m\mathrm{d}x}{\int_0^1 x^m(1-x)^m\mathrm{d}x}.\label{eq:phieq}
\end{equation}
This interpolation was originally suggested in ref.~\cite{rezakhani:adiabaticexponential} and is chosen because it conveniently guarantees that the first $m$ derivatives of $\mathcal{H}(s)$ are zero at $s=0$ and 
$s=1$.  Additionally, in the $m=0$ case it gives the linear interpolation $\phi(s)=s$ used in Fig.~\ref{fig:adiabaticsubsequence}. 

Fig.~\ref{fig:boundarycancel} demonstrates the improvements that arise from combining our results with those taken from refs.~\cite{rezakhani:adiabaticexponential,lidar:exponentialtime}.  It is notable to see that the $m=0$ data in Fig.~\ref{fig:boundarycancel} (a) nearly coincides
with that for $m=1$ in Fig.~\ref{fig:boundarycancel} (b) for sufficiently large $T$. Similarly, the $m=1$ data in Fig.~\ref{fig:boundarycancel} (a) corresponds to the $m=2$ data in~\ref{fig:boundarycancel} (b) in the same limit. This shows that our technique can be used to improve the overall accuracy of boundary cancellation techniques \emph{without} compromising the error-scaling for short $T$.


\begin{figure}[t!]
\includegraphics[width=\linewidth]{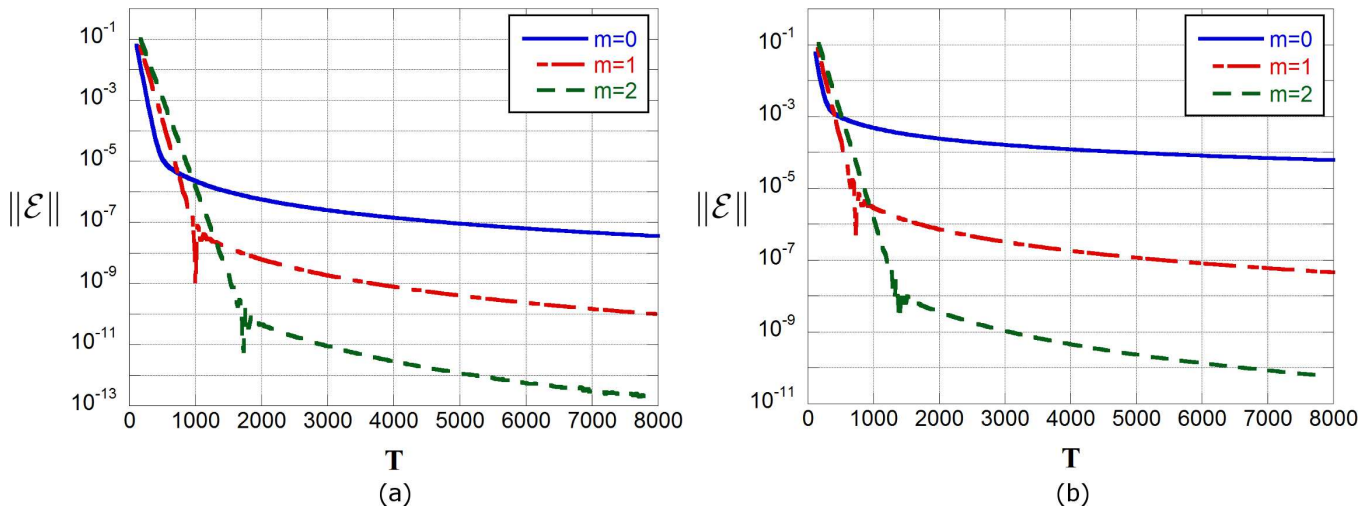}\caption{This figure shows that existing boundary cancellation methods can be augmented with our boundary cancellation method to achieve even higher-order error-scaling for a search Hamiltonian with $N=16$ and $\phi(s)$ chosen as in eq.~\eqref{eq:phieq}. Fig.~\ref{fig:boundarycancel} (a) is a plot of the error at the times when our theory predicts improved error-scaling (i.e., even $n$), whereas Fig.~\ref{fig:boundarycancel} (b) displays the times when the errors are predicted to be maximized (i.e., odd $n$).}\label{fig:boundarycancel}
\end{figure}
\section{Two-Qubit Gate.}\label{sec:gate}
Our technique naturally lends itself to Hamiltonians that couple the ground state to only one excited state, such as the Search Hamiltonian given in eq.~\eqref{eq:searchHam}. If the total error $\|\mathcal{E}\|$ is dominated by several transitions, this technique can still be adapted to approximately cancel multiple transitions simultaneously. To demonstrate, we show how to optimize the fidelity of an adiabatic two-qubit logic gate without decreasing its speed. Similar improvements were reported previously \cite{Charron:2002}, without a broadly-applicable underlying theory or error bounds.

We apply of our method to an exchange-based two-qubit operation designed for neutral atom QIP \cite{Anderlini:2007,Hayes:2007,Babcock:2008, Stock:2008}. This operation exploits identical particle exchange to generate a partial ``swap'' operation between qubits stored in nuclear spin \cite{Hayes:2007} or valence electronic states \cite{Stock:2008} of optically trapped atoms. The gate generates a relative phase of $e^{-i\alpha}$ between the symmetric and antisymmetric components of the particles' vibrational degrees of freedom. The phase difference is then transferred to the respective components of the two-qubit subspace $\big\{|ij\rangle: i,j\in\{0,1\}\big\}$. This produces an operation that (with single-qubit rotations) is locally equivalent to a tunable entangling controlled-phase gate $e^{-2i\alpha|11\rangle\langle11|}$\cite{Stock:2008}. 

Following previous work~\cite{Babcock:2008, Stock:2008}, we examine a simple Hamiltonian governing two identical particles confined to one dimension and trapped by pair of moving potential wells.  The Hamiltonian for particles 1 and 2 
is given by

\begin{equation}
\mathcal{H}(x_1,x_2,p_1,p_2,s) = \mathcal{H}(x_1,p_1,s) + \mathcal{H}(x_2,p_2,s) +2a\omega_\bot\delta(x_1\!-\!x_2), 
\label{eq:twoatomHam}\end{equation}
for $\mathcal{H}(x,p,s)= p^2/2m+V(x+(s-\frac{1}{2})d)+V(x-(s-\slantfrac{1}{2})d)$, where $x$ and $p$ are the position and momentum of a particle of mass $m$. The potential $V(x)=-V_\text{o}\exp(-x^2/2\sigma^2)$ describes a 1D Gaussian trap of depth $V_\text{o}$ and variance $\sigma^2$. Traps are initially separated by a distance $d=3\sigma$. We consider a 1D s-wave scattering interaction, with scattering length $a\omega_\bot=3\sigma$ and transverse confinement frequency $\omega_\bot$~\cite{Calarco:2000}. 
As eq.~\eqref{eq:twoatomHam} is symmetric, transitions between symmetric and antisymmetric states are forbidden, and each symmetry subspace evolves independently. 

We diagonalized eq.~\eqref{eq:twoatomHam} over the range $0\le s\le0.5$ at $\Delta s=1/1200$ intervals. We then used a spline fitting to integrate eq.~\eqref{eq:searchDestruct}, obtaining numerical estimates $T$ of the ideal $T_{n,\nu}$. The quality of initial approximations were then improved using the relationship $|T_{n,\nu}- T|\approx T/\Delta n$, where $\Delta n$ measures the beat frequency between $T^{-1}$ and $T_{n,\nu}^{-1}$ (e.g., the distance between cusps on inset, Fig.~\ref{fig:gate}). More sophisticated model Hamiltonians may be solved using more advanced numerical techniques and empirically refined in the same manner.

We numerically integrated the Shr\"odinger equation to obtain system dynamics of durations $\{T_{n,5}\}$, explicitly generating sets of wave functions $\{|\psi^+_n(s)\rangle\}$ and $\{|\psi^-_n(s)\rangle\}$ for two distinct initial states: the symmetric ground state $|\psi^+_n(0)\rangle=|0(0)\rangle$ and  the antisymmetric (effective) ground state $|\psi^-_n(0)\rangle=|1(0)\rangle$. We chose $T_{n,5}$ because $\ket{5(s)}$ is the first eigenstate that significantly couples to $\ket{0(s)}$.  This transition is dominant because the
  $0\!\leftrightarrow\!1$, $0\!\leftrightarrow\!2$, and $0\!\leftrightarrow\!3$
  transitions are forbidden, and the $0\!\leftrightarrow\!4$ coupling is
  weak. We define $|\!\!\:\langle\psi^{\!\pm\;}_n\!\!|\nu\rangle\!\!\:|=|\langle\psi^{\pm}_n(1)|\nu(1)\rangle|$.

The error probabilities are improved by nearly three orders of magnitude over the bound set by eq.~\eqref{eq:standardcriterion} by applying our technique to this system (Table~\ref{mytable}). This corresponds to a tenfold increase in gate speed (given a maximum error rate of $10^{-4}$), for the linear motion described by eq. \eqref{eq:twoatomHam}. Greater improvements could be achieved by choosing $\mathcal{H}(x,p,s)$ or $s(t)$ to satisfy eq.~\eqref{eq:searchDestruct} for more transitions simultaneously and with better synchronization.

Partial swap operations have been experimentally demonstrated using neutral atoms in a double-well optical lattice, but the adiabatic requirement limits gate times ($\sim \!\!4\,\text{ms}$ for high fidelity operation \cite{Anderlini:2007}). Our technique thus affords a significant advancement to inherently slow gates of this kind. Furthermore, because the phase $\alpha$ scales with $T$ (see Table~\ref{mytable}), the precision necessary for accurate gate operation is itself comparable to that needed to implement our phase cancellation technique on an atomic quantum logic gate.

We have numerically demonstrated that error in the adiabatic approximation can be reduced for an experimentally relevant model of a quantum gate.  An important remaining issue is whether the experimental uncertainties required to observe error reductions are reasonable for this model system. By first-order Taylor expansion of eq.~\eqref{eq:firstorder1}, we find that if
\begin{align}
\Delta S_5\left/ \left( \frac{\beta_{5}(0)}{\gap_5(0)}\right)\right.< 33\% \quad \text{ and } \quad \frac{\Delta G_5}{\int_0^1\gap_5(s)\mathrm d s}=\frac{\Delta T_5}{T_{460,5}}< 0.02\%,
\end{align}
then the \emph{observed} transition amplitude at $T\approx T_{460,5}$ will be less than half of that at $T=T_{459,5}$. These modest requirements imply that our $m=0$ method may be rapidly incorporated into present-day or near-future atom-based QIP experiments.  Such an experiment would also provide a highly sensitive test of the validity of the adiabatic approximation in open quantum systems.

\begin{figure*}[t]
\begin{minipage}{0.48\linewidth}
\center{\includegraphics[width=\textwidth]{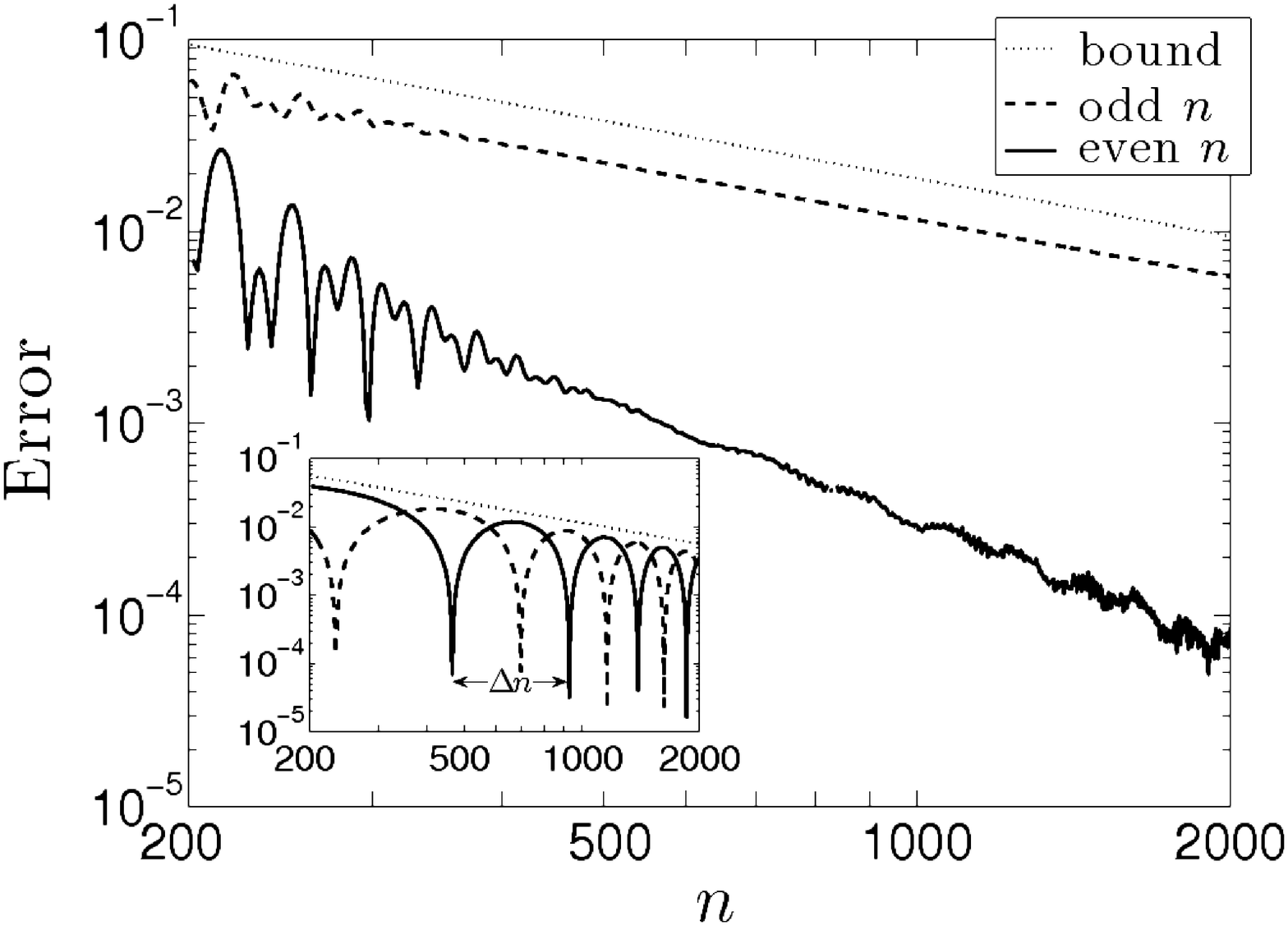}}
\caption{Transition amplitudes and bounds for $T=T_{n,5}$, over $200\le n \le 2000$. Main figure shows $|\!\!\:\langle\psi^{\!+\;}_n\!\!|5\rangle\!\!\:|\approx|\mathcal{E}_5|$ for even $n$ (solid) and odd $n$ (dashed), which are bounded by $\max_s[2\left\|\frac{\mathrm{d}}{\mathrm{d}s} H(x,p,s)\right\|/(E_5(s)\!\!-\!E_0(s))^2]$ (dotted). Inset shows $|\!\!\;\langle\psi^{\!-\;}_n\!\!|6\rangle\!\!\;|$ bounded by $\max_s[2\left\|\frac{\mathrm{d}}{\mathrm{d}s} H(x,p,s)\right\|/(E_7(s)\!\!-\!\!E_1(s))^2]$.}\label{fig:gate}

\end{minipage}
\hspace{0.5cm}
\begin{minipage}{0.48\linewidth}
\flushleft

\vspace{0.09in}
\begin{tabular}{@{\hspace{-0.01cm}} c @{\hspace{0.2in}} r@{.}l @{\hspace{0.12in}} r@{.}l @{\hspace{0.12in}} r@{.}l @{\hspace{0.18in}} r@{.}l @{\hspace{0.20in}} r@{.}l @{\hspace{0.12in}} r@{.}l}
\hline
Run & \multicolumn{10}{c}{$\!\!\!\!$Error Probabilities$\;(\times\;10^{-4})$}  &  \multicolumn{2}{r}{Phase} \\
$n$ &   \multicolumn{2}{l}{$\!\!\!|\!\!\:\langle\psi^{\!+\;}_n\!\!|5\rangle\!\!\:|^2$} & \multicolumn{2}{l}{$\;\:\!|\!\!\:\langle\psi^{\!-}_n\!\!\:|6\rangle\!\!\:|^2$}  &  \multicolumn{2}{l}{$\,|\:\!\!\langle\psi^{\!-}_n\!\!\:|7\rangle\!\!\:|^2$} & \multicolumn{2}{l}{$\!\!\|\mathcal{E}^+\|^2$} & \multicolumn{2}{l}{$\!\!\|\mathcal{E}^-\|^2$} &  \multicolumn{2}{l}{$\quad\;\,\alpha$}\\
\hline
456 & $0$&$024$ & $0$&$012$   & $0$&$245$  & $0$&$988$ & $0$&$535$ & $ 1$&$645$\\
458 & $0$&$022$ & $0$&$007$   & $0$&$180$  & $0$&$771$ & $0$&$433$ & $-0$&$186$\\
460 & $0$&$021$ & $0$&$003$   & $0$&$124$  & $0$&$648$ & $0$&$316$ & $ 4$&$266$\\
462 & $0$&$021$ & $0$&$001$   & $0$&$078$  & $0$&$764$ & $0$&$275$ & $-3$&$849$\\
464 & $0$&$023$ & $<\!0$&$0001$ & $0$&$043$  & $0$&$980$ & $0$&$249$ & $ 0$&$603$\\
466 & $0$&$023$ & $<\!0$&$0003$  & $0$&$018$  & $1$&$010$ & $1$&$201$ & $-1$&$228$\\
468 & $0$&$023$ & $0$&$002$   & $0$&$003$  & $0$&$925$ & $0$&$292$ & $-3$&$059$\\
470 & $0$&$023$ & $0$&$004$   & $\!<\!0$&$0001$ & $0$&$763$ & $0$&$452$ & $1$&$392$\\
472 & $0$&$022$ & $0$&$007$   & $0$&$006$  & $0$&$721$ & $0$&$502$ & $-0$&$438$\\
474 & $0$&$021$ & $0$&$012$   & $0$&$021$  & $0$&$803$ & $0$&$654$ & $ 4$&$014$\\
\hline
\end{tabular}

\renewcommand{\figurename}{TABLE}
\renewcommand{\thefigure}{\Roman{figure}}
\setcounter{figure}{0}
\caption[labelformat=empty,labelsep=colon]{Error probabilities and the phase gap $\alpha$ (radians) obtained from simulation runs $\{T_{n,5}\}$ for $456 \le n \le 474$. For these times, local minima of $|\!\!\;\langle\psi^{\!+\;}_n\!\!|5\rangle\!\!\;|$ roughly match those of $|\!\!\:\langle\psi^{\!-}_n\!\!\:|6\rangle\!\!\:|$ (inset, Fig.~\ref{fig:gate}) and $|\:\!\!\langle\psi^{\!-}_n\!\!\:|7\rangle\!\!\:|$. We denote total errors as $\|\mathcal{E}^+\|^2 = 1-|\!\!\:\langle\psi^{\!+\;}_n\!\!|0\rangle\!\!\:|^2$ and $\|\mathcal{E}^-\|^2 = 1-|\!\!\:\langle\psi^{\!-\;}_n\!\!|1\rangle\!\!\:|^2$. Eq.~\eqref{eq:standardcriterion} predicts $\|\mathcal{E}^+\|^2\le0.046$ and $\|\mathcal{E}^-\|^2\le0.62\times10^{-3}$ at $n=460$.
} \label{mytable}

\end{minipage}

\end{figure*}

\section{Conclusion}
We have presented a new technique for improving the fidelity of adiabatic transport. Our technique exploits an adiabatic phase cancellation effect that occurs at certain evolution times to produce improved error-scaling.  In addition, our method applies directly to a host of experimentally relevant physical systems, often without modification to the adiabatic path $s(t)$.
Our technique can also be used to improve the accuracy of existing boundary cancellation techniques, providing improved scaling over those methods when an easily-satisfiable symmetry condition~\eqref{eq:condition} is met. 
We show that these ``augmented'' boundary cancellation techniques
can provide unsurpassed accuracy, requiring comparably precise
control over the Hamiltonian to achieve high-order error-scaling. Consequently, our
work reveals that precision (in addition to energy and time) is a
subtle and important resource to consider when devising algorithms and experiments that utilize adiabatic state transfer.

We have illustrated these claims using numerical examples of QIP applications.
We numerically demonstrated the use of augmented boundary cancellation methods for $m=0,1,2$ for
an adiabatic search algorithm. We also optimized a simple adiabatic quantum logic gate using our $m=0$ method.  In that case we also performed an error analysis and found that the error tolerances needed to apply the method are experimentally reasonable.

Our results open several interesting avenues of further inquiry.  
We have shown that our technique can be used
to improve the accuracy of some local adiabatic evolutions, but
it would be interesting to see if further improvements can be obtained
by using our method in concert with more sophisticated adiabatic optimization
methods such as the one given in ref.~\cite{rezakhani:geodesic}.  In addition, determining
the error tolerances for small deviations along the adiabatic path would
be an important step towards fully characterizing precision as
a resource for adiabatic processes. Our preliminary estimates suggest that
it may be possible to observe error reductions for atom-based quantum logic using optical dipole traps,
but other experimental setups may also be well-suited to
study this effect, such as nuclear magnetic resonance (NMR) systems.
Such experiments would not only be interesting as a test of
the viability of augmented boundary cancellation methods
as an error-reduction strategy, but would also provide a highly
sensitive test of the limits of the adiabatic approximation itself.

\appendix
\section{Proof of Eq.~\ref{eq:generalboundary}\label{appendix:1}}
In Section~\ref{sec:theory} we claimed that phase cancellation can be used to accelerate the convergence of boundary cancellation techniques.  Specifically, we claimed that our augmented boundary cancellation methods reduce $|\mathcal{E}_\nu|$ from order $\mathcal{O}(T^{-m-1})$ to $\mathcal{O}(T^{-m-2})$.  We will now justify why this is the case.

Using the path-integral representation of the time-evolution operator presented in~\cite{farhi:adiabaticpaths,mackenzie:adiabatic,cheung:adiabatic}
we have that

\begin{equation}
|\mathcal{E}_\nu(1)|= \left\|\int_0^1 \beta_{\nu,0}(s)e^{-i\int_0^s \gap_{\nu}(\xi)\mathrm{d}\xi T} \mathrm{d}s+\sum_{\mu}\int_0^1 \beta_{\nu,\mu}(s)e^{-i\int_0^s \gap_{\nu,\mu}(\xi)\mathrm{d}\xi T}\int_0^s \beta_{\mu,0}e^{-i\int_0^{s_2} \gap_{\mu}(\xi) \mathrm{d}\xi T} \mathrm{d}s_2\mathrm{d}s + \cdots\right\|.\label{eq:pathint}
\end{equation}
where $\beta_{\nu,\mu}$ is defined for any $\nu$ and $\mu$ in the set $\{0,\ldots,N-1\}$ by
\begin{align}
\beta_{\nu,\mu}(s)
= \left\{\begin{array}{cl}0& \textrm{if $E_{\nu}(s)=E_\mu(s),$}\\ \frac{\bra{\nu(s)}\dot{\mathcal{H}}(s) \ket{\mu(s)}}{E_\nu(s)-E_\mu(s)} & \textrm{otherwise.}\end{array}\right.\label{eq:betadef2}
\end{align}
We analyze the series under the assumption that the first $m$ derivatives of the Hamiltonian are zero at the boundaries $s=0,1$.  Using integration by parts,
we find that
\begin{equation}
\int_0^1 \beta_{\nu,0}(s)e^{-i\int_0^s \gap_{\nu}(\xi)\mathrm{d}\xi T} \mathrm{d}s= \left.\frac{\bra{\nu(s)} \mathcal{H}^{(1)}(s)\ket{0(s})}{-i\gap_{\nu}^2(s)T}e^{-i\int_0^s \gap_{\nu}(s)\mathrm{d}s T}\right|_0^1-\int_0^1 \left(\frac{\partial}{\partial_s} \frac{\beta_{\nu,0}(s)}{-i\gap_{\nu}(s)T}\right)e^{-i\int_0^s \gap_{\nu}(\xi)\mathrm{d}\xi T} \mathrm{d}s\label{eq:genintparts1}
\end{equation}
Then, using the fact that $\mathcal{H}^{(1)}(0)=\mathcal{H}^{(1)}(1)=0$, the first term on the right side of eq.~\eqref{eq:genintparts1} is zero.  Evaluating the second term using integration by parts, we obtain
\begin{equation}
- \left.\left(\frac{\partial}{\partial_s} \frac{\bra{\nu(s)}\mathcal{H}^{(1)}(s)\ket{0(s)}}{-i\gap_{\nu}(s)^2T}\right)e^{-i\int_0^s \gap_{\nu}(\xi)\mathrm{d}\xi T}\right|_0^1+\int_0^1 \left(\frac{\partial}{\partial_s}\frac{1}{\gap_{\nu}(s)T}\left(\frac{\partial}{\partial_s} \frac{\beta_{\nu,0}(s)}{-i\gap_{\nu}(s)T}\right)\right)e^{-i\int_0^s \gap_{\nu}(\xi)\mathrm{d}\xi T} \mathrm{d}s.
\end{equation}
As before, the first term in this expression is zero because the first two derivatives of the Hamiltonian are zero. We then continue by this reasoning, applying integration by parts $m+1$ times.  Then after dropping the first $m$ derivatives of the states, Hamiltonian, and energy gaps at $s=0,1$, we find
\begin{align}
&\left|\int_0^1 \beta_{\nu,0}(s)e^{-i\int_0^s \gap_{\nu}(\xi)\mathrm{d}\xi T} \mathrm{d}s\nonumber\right|\\
&\qquad\qquad=\left|\frac{1}{T^{m+1}}\left(\frac{\bra{\nu(1)}\mathcal{H}^{(m+1)}(1)\ket{0(1)}e^{-i\int_0^1\gap_{\nu}(s)\mathrm{d}s T}}{\gap_{\nu}(1)^{m+2}}-\frac{\bra{\nu(0)}\mathcal{H}^{(m+1)}(0)\ket{0(0)}}{\gap_{\nu}(0)^{m+2}}\right)\right|+\mathcal{O}(1/T^{m+2}).\label{eq:higherorderbdy}
\end{align}
We then see that the symmetry condition in eq.~\eqref{eq:condition} implies that if $T=T_{n,\nu}$ then the first term in the expansion in eq.~\eqref{eq:genintparts1} is $\mathcal{O}(1/T^{m+1})$.  The result of eq.~\eqref{eq:generalboundary} then holds if the remaining terms in eq.~\eqref{eq:genintparts1} are asymptotically negligible.

Turning our attention the remaining path-integrals in eq.~\eqref{eq:genintparts1}, we find that all of the remaining terms are $\mathcal{O}(1/T^{m+2})$.  This is because these terms involve contain multiple products of $\beta_{\mu,\nu}$.  Therefore, if we perform integration
by parts $m+1$ times on the outermost integral, then the term involving $\mathcal{H}^{(m+1)}$ becomes multiplied by at least one $\beta_{\mu,\nu}$ term, which is zero on the boundary by definition. Therefore, no non-zero terms appear in the expansion of these integrals to $\mathcal{O}(1/T^{m+2})$. Hence, the first term in eq.~\eqref{eq:genintparts1} is asymptotically dominant as anticipated~\cite{mackenzie:adiabatic,cheung:adiabatic}.  Since the first term is asymptotically dominant and also of order $\mathcal{O}(1/T^{m+2})$ given
 $T=T_{n,\nu}$, the augmented boundary cancellation technique proposed in Sec.~\ref{sec:theory} combines with existing methods.

\section{Error-Robustness of Augmented Boundary Cancellation Methods}\label{appendix:2}
In Section~\ref{sec:error1} we claimed without proof that if the uncertainty in the $p^{\rm th}$ derivative of $\mathcal{H}(s)$ is $\mathcal{O}(T^{-m-2+p})$ for all $p=1,\ldots,m$, then that derivative can safely be assumed to be negligible.  We prove this now by demonstrating that the leading order terms involving $\mathcal{H}^{(p)}(0)$  or $\mathcal{H}^{(p)}(1)$ for $p=1,\ldots,m$ are of order~$\mathcal{O}(T^{-m-2})$ under this assumption.  

We begin by assuming that, for some $q$, $\mathcal{H}^{(q)}(s)$ is non-zero at the boundaries $s=0,1$ and that all lower derivatives are negligible there.  Following the argument put forward in Appendix~\ref{appendix:1}, the lowest order term that appears after applying integration by parts $q$ times to eq.~\eqref{eq:pathint} is
\begin{align}
\left|\frac{1}{T^{q}}\left(\frac{\bra{\nu(1)}\mathcal{H}^{(q)}(1)\ket{0(1)}e^{-i\int_0^1\gap_{\nu}(s)\mathrm{d}s T}}{\gap_{\nu}(1)^{q+1}}-\frac{\bra{\nu(0)}\mathcal{H}^{(q)}(0)\ket{0(0)}}{\gap_{\nu}(0)^{q+1}}\right)\right|.\label{eq:higherorderbdy2}
\end{align}

If $\mathcal{H}^{(q)}(1)$ and  $\mathcal{H}^{(q)}(0)$ are both of order $\mathcal{O}(T^{-m-2+q})$, then the term~\eqref{eq:higherorderbdy2} is reduced to order $\mathcal{O}(T^{-m-2})$.  As argued in Appendix~\ref{appendix:1}, other terms that appear in the perturbative series after repeated integrations by parts are asymptotically smaller than this term and therefore do not affect the error-scaling.  Thus, it is sufficient to render errors in the $q^{\rm th}$ derivative of $\mathcal{H}(s)$ negligible by taking them to be $\mathcal{O}(T^{-m-2+q})$.

By the same reasoning, if the uncertainty in the $p^{\rm th}$ derivative of $\mathcal{H}(s)$ is $\mathcal{O}(T^{-m-2+p})$ for all $p=1,\ldots,m$, then the total contribution of derivative errors is $\mathcal{O}(T^{-m-2})$ given that $m\in \mathcal{O}(1)$.  This implies that augmented boundary cancellation methods are robust to derivative errors given that $m$ is a fixed integer.  This result also trivially implies that existing boundary cancellation methods are robust to derivative errors under the same circumstances.

If $m$ is not bounded from above by a constant then this analysis fails because the previous analysis ignored multiplicative factors of $m$ that appear in the analysis.  Such terms could make the neglected higher-order derivative terms much larger if $m$ is an increasing function of $T$.  This means that if we wish to achieve exponential error-scaling by taking $m\in \Theta(T/\log(T))$, then the tolerance for derivative errors must shrink even further from the already exponentially small error tolerances obtained by substituting $m\in \Theta(T/\log(T))$ into $\mathcal{O}(T^{-m-2+p})$ for fixed $p$.  We conclude that boundary cancellation methods that exhibit exponential error-scaling are not robust to derivative errors.

\begin{acknowledgements}
We wish to thank Emily Pritchett, Mark Raizen, and Barry Sanders for helpful discussions. This work was supported by NSERC, AIF, iCORE, MITACS research network, General Dynamics Canada and  USARO.
\end{acknowledgements}

\end{document}